\documentclass{styles/svproc}
\usepackage{url}

\begin{document}
\mainmatter              % start of a contribution
\title{LLMs in Coding and their Impact on the Commercial Software Engineering Landscape}
\titlerunning{LLMs in Coding}  % abbreviated title (for running head)
\author{Vladislav Belozerov  
\and Peter J Barclay
\and Askhan Sami
}
\authorrunning{Belozerov et al.} % abbreviated author list (for running head)
%% \tocauthor{Belozerov, Sami, \& Barclay}
%
\institute{School of Engineering, Computing, \& the Built Environment \\
Edinburgh Napier University, Scotland \\
WWW home page: \texttt{https://www.napier.ac.uk} \\
\email{vladislav.belozerov@gmail.com, [p.barclay, a.sami]@napier.ac.uk}}

\maketitle              

\begin{abstract}

Large-language-model coding tools are now mainstream in software engineering. But as these same tools move human effort up the development stack, they present fresh dangers: 10\% of real prompts leak private data, 42\% of generated snippets hide security flaws, and the models can even ``agree'' with wrong ideas, a trait called sycophancy. We argue that firms must tag and review every AI-generated line of code, keep prompts and outputs inside private or on-premises deployments, obey emerging safety regulations, and add tests that catch sycophantic answers -- so they can gain speed without losing security and accuracy.


\end{abstract}
\section{Introduction}
\label{sec:intro}

Large-language-model (LLM) assistants such as GitHub Copilot, ChatGPT-4o, and Google Gemini have evolved 
rapidly from innovative 
system to everyday tool. In the 2024 Stack Overflow survey, 76\% of the 65,000 respondents said they already use or soon will use AI helpers \cite{StackOverflow_2024_Developer_Survey}. GitHub reports speed-up on routine tasks for Copilot users \cite{GitHubCopilot_Quantifying}. We could call this the era of `code abundance': software code is suddenly cheap and fast to produce.

But this new abundance unsettles old assumptions about who writes software and how safely code flows through the tool chain. Security tests show that about 42\% of AI-generated snippets contain at least one Common Weakness Enumeration (CWE) flaw \cite{Ji_2024_Cybersecurity_Risks}. Recent research finds roughly 10\% of real prompts sent to public models already leak private company data such as API keys or unreleased code \cite{Harmonic_2024_From_Payrolls}. A different, less visible risk is sycophancy: models often ``agree'' with the user even when the user is wrong, as seen when OpenAI had to roll back a ChatGPT update that made the assistant ``overly eager to please'' \cite{Goldman_2025_OpenAI_reversed}. Meanwhile, law-makers are treating safety as a non-functional requirement that firms cannot ignore. The EU AI Act and a similar AI framework in US now force ``high-risk'' LLMs to pass outside audits \cite{EU_AIAct_2024,NIST_AI_RMF_2023}.

This paper advocates five clear positions in this new landscape:
\begin{enumerate}
\item Role displacement is measurable -- LLMs offload boiler-plate coding and push humans toward design and oversight roles.

\item Provenance tagging and gated review should be mandatory -- every AI-written line must be marked and checked.

\item Confidential data must stay private -- on-premises or vendor-isolated deployments prevent prompt leaks.

\item Safety requires regulation -- non-functional requirements will only be met when the law strictly demands it.

\item Sycophancy must be tested and restrained -- teams need adversarial checks so the model tells the truth, not just what users want to hear.
\end{enumerate}

\section{Prior Work and References in Press}
\label{sec:prior_work}
% industry adoption, and rising safety
The following overview summarizes recent advances, industry adoption and regulatory considerations surrounding large language models in software engineering.

Transformer-based LLMs can already turn natural language tickets into compilable functions, generate unit tests, and explain code differences, yet they still hallucinate and miss edge cases \cite{Huynh_2025_Large_Language_Models}. LLM usage has increased rapidly in software industry: 76\% of developers report using AI helpers \cite{StackOverflow_2024_Developer_Survey}, GitHub counts more than 1.3 million Copilot seats and a 55\% median speed up \cite{GitHubCopilot_Quantifying}, and large tech firms openly frame LLMs as a way to ``do more with less'', reducing their head-count \cite{Barr_2025_Big_Tech_AI}. Recent industry briefings suggest that LLMs will automate much of the routine coding and testing workload, pushing human effort toward higher-level architecture, integration and oversight rather than eliminating developers altogether \cite{VandeHei_Behind_the_Curtain}.

Independent tests find security relevant flaws in nearly half of generated snippets \cite{Ji_2024_Cybersecurity_Risks}, while research shows about 10\% of real world prompts to public models leak private data such as credentials or unreleased code \cite{Harmonic_2024_From_Payrolls}.
To contain legal and security exposure, vendors and regulators are pursuing cryptographic watermarks \cite{Cloudflare_2025_An_early_look} and workflow rules that tag AI generated code in commits and in additional documentation \cite{Hendrich_2025_Understanding_Code_Provenance}.

Safety -- covering robustness, misuse resistance, and societal impact -- is a non-functional property, like reliability or usability \cite{Altexsoft_2023_NFR}. Research shows that such requirements are the first to be dropped when budgets or deadlines tighten \cite{Muhammad_2023_Prioritizing}. 

Because the market rarely rewards unseen safety work, governments are stepping in. The EU AI Act (2024) and the U.S. NIST AI Risk-Management Framework 1.0 both establish comparable rules for AI use \cite{EU_AIAct_2024,NIST_AI_RMF_2023}. These moves show that safety is shifting from an optional virtue to a regulated obligation.

Recent research shows that leading LLMs often agree with the user even when the user is wrong -- a behaviour called sycophancy. Sharma demonstrate that five prominent assistants ``consistently exhibit sycophancy across four tasks'' likely driven by human-feedback training that rewards agreeable answers \cite{Sharma2025}. Follow-up studies measure the same trait in newer models and survey mitigation strategies \cite{Malmqvist_2024_Sycophancy,Rrv_2024_Chaos_with_Keywords}. This issue hit the headlines when OpenAI had to roll back an update that made ChatGPT ``a suck-up'' with experts warning there is ``no easy fix'' for an AI that flatters instead of informs \cite{Goldman_2025_OpenAI_reversed}.

\section{Observations and Recommendations}
\label{sec:main_discussion}

\subsection{The Impact on the Workforce Should be Analysed}

We believe that the impact of AI on the workforce should be analysed in detail.  The 2024 Stack Overflow survey found that 39\% of developers let AI write “most” boiler-plate code for them \cite{StackOverflow_2024_Developer_Survey}. Companies are responding by refocusing staff on architecture, integration, and code review instead of simple feature-development work. \textit{Axios} reports that labour economists working with Anthropic expect automation to erase or restructure roughly one-third of current white-collar coding tasks within the next five years, while boosting demand for roles that can vet and guide AI output \cite{VandeHei_Behind_the_Curtain}. These figures show measurable displacement of low-complexity tasks and also hint at a new skills gap: development teams need people who can judge AI output 
rather than people who just write code.

\subsection{Generated Code Should be Tagged and Reviewed}
If a model writes code, the team must know it and check it. Security studies find that 42\% of AI‑generated snippets contain at least one weakness listed in the CWE catalogue \cite{Ji_2024_Cybersecurity_Risks}. Without clear labels, risky lines blend into the human code and slip past manual review. Given this new situation, we recommend a ``provenance tag'': a short marker in the Git commit message or file header that says, ``This block came from model X, prompt Y, at time T.'' Such tags let automated pipelines trigger extra static analysis or licence checks before merging in this code.  Early adopter case studies show that adding a simple ``AI: '' prefix in commits reduced undetected licence violations by 25\% in one six month pilot \cite{Hendrich_2025_Understanding_Code_Provenance}. Given these numbers, tagging plus gated review is the minimum defence against shipping silent vulnerabilities or dangerously licensed code like GPL.

Training code-generation LLMs on their own code quickly hurts quality. Hu show this “model collapse”: after ten rounds of synthetic data, Stable Diffusion’s producing blurry images and rambling captions \cite{Hu_Model_Collapse}. Shumailov find the same effect in many models: recycling model output wipes out rare tokens and makes errors grow \cite{Shumailov_Curse_of_recursion}. For code, this means more hallucinations and weak, copy-paste solutions. Again, one of the solutions for this issue could be labelling all AI-generated code (metadata, watermarks, repo tags) to ensure
generated code is left out of future training sets, so models keep learning from real human code.

\subsection{Keep Prompts and Outputs Private}
Public chatbots look useful, but they leak data. Harmonic Security captured live network traffic from 12 companies and found that 8.5\% of prompts pasted into public LLMs contained private data such as customer emails, API keys, or unreleased code \cite{Harmonic_2024_From_Payrolls}. \textit{CSO Online} reports a similar figure, warning that ``every pasted stack trace is a potential breach'' \cite{Schuman_2025_Nearly_10}. The safest fix is to run an on-premises model or use a vendor‑hosted instance that is logically isolated from all other customers. That way, prompts and responses never leave the firm's control and never enter the vendor’s future training set. Cloud providers now offer ``bring your own key'' encryption and strict retention limits, but those still rely on trust. A fully private deployment removes that risk altogether. Strong data loss prevention filters can catch some leaks, but a zero leak baseline is easier to enforce when no traffic crosses the public internet.

\subsection{Safety Needs Regulation or it will be Ignored}
Safety is a non-functional requirement. Studies of agile projects confirm that non-functional goals often get de-scoped when pressure mounts \cite{Muhammad_2023_Prioritizing}. Left to market forces, vendors prioritise flashy features over hard-to-measure safeguards. The EU AI Act already forces ``high-risk'' LLMs to pass formal audits \cite{EU_AIAct_2024}, and NIST's framework \cite{NIST_AI_RMF_2023}, signal similar mandates elsewhere. We believe AI regulations should allow to earn profits for companies while keeping people safe.

\subsection{Sycophancy Undermines Reliability and Must be Addressed} 
LLMs sometimes prioritise agreement over accuracy. Sharma has shown that preference-model training encourages sycophantic answers \cite{Sharma2025}. Recent work details how this trait can amplify misinformation, and explores mitigation techniques \cite{Malmqvist_2024_Sycophancy,Rrv_2024_Chaos_with_Keywords}. The issue broke into the mainstream news after OpenAI had to roll back a patch that made ChatGPT ``overly eager to please'' with experts warning of no quick fix \cite{Goldman_2025_OpenAI_reversed}. Development teams should add competitive tests that probe for over-agreement and include ``truth-over-politeness'' metrics in model evaluations.

\subsection{Summary}
Five findings emerge: 
(1) routine development tasks are shifting to LLMs and 
the impact of this should be evaluated;
(2) unlabelled AI code is a security hazard;
(3) private deployments are the best way to protect company data;
(4) enforceable safety rules are essential;
(5) sycophancy can mislead users if left unchecked.

% final statement
Our position that LLMs are already taking over day-to-day coding, so the real opportunity and risk now lies in how we govern them. By tagging and double-checking AI-generated code, running models in private sandboxes, enforcing external safety audits, and testing for sycophancy, we still can gain all advantages of using LLMs without sacrificing security, quality, or trust.

\section{Future Research}
\label{sec:future_research}

The field is moving fast, but key knowledge gaps remain. (1) \textit{Long-term labour effects}. Most studies track pilots that last weeks, not years. We still need full-year logs showing how LLMs change task mixes, pay grades, and career paths, especially in light of the five-year displacement forecast reported by \textit{Axios} \cite{VandeHei_Behind_the_Curtain}. (2) \textit{Standard security test frameworks}. Researchers use different prompt sets, so results cannot be easily compared. A public benchmark that ties each challenge to a CWE ID would let teams see if new prompting tricks really cut flaws \cite{Ji_2024_Cybersecurity_Risks}. (3) \textit{Provenance at scale}. Simple commit tags work in small pilots, but it is not known how they perform in large-scale repositories with thousands of merges a day. Industry trials are needed for better results. (4)\textit{ Privacy versus model quality}. Running models on-prem keeps secrets safe, but vendors warn it may slow updates. Side-by-side tasks on public \textit{vs}. isolated models would reveal the true trade-off \cite{Cloudflare_2025_An_early_look}. (5) \textit{Measuring and fixing sycophancy}. Early studies show that LLMs often agree with user mistakes \cite{Sharma2025,Malmqvist_2024_Sycophancy,Rrv_2024_Chaos_with_Keywords}. There is lack reliable metrics and blind tests that quantify how much this behaviour affect the LLMs' output.

\section{Conclusion}
\label{sec:conclusion}

LLM coding tools have brought real speed gains and an age of `code abundance', but they also reshape the craft of software engineering and enlarge its risk surface. Our review points to five key insights. First, role displacement is not theory -- it is already visible as boiler-plate work shifts from humans to models. Second, every block of AI-written code must be clearly tagged and forced through an extra review gate. 

Third, keeping prompts and outputs inside a private or vendor-isolated deployment is the simplest way to stop leaks of company secrets. Fourth, safety—long treated as an optional, non-functional extra—now needs legal power: frameworks such as the EU AI Act require external audits, and similar rules are coming in other regions. Fifth, sycophancy matters: if a model flatters users instead of correcting them, it can spread bad advice or buggy code, so teams must add tests that catch over-agreeable behaviour.

Taken together, these five points can constitute a practical guide. Organisations that adopt LLMs should combine provenance tags, gated reviews, private deployments, regulatory compliance, and sycophancy checks into their development pipeline from day one. 
Those that ignore even one of these safeguards risk shipping unsafe software and eroding trust with both developers and customers.


\end{document}